\documentclass[11pt]{article}
\usepackage[margin=1in]{geometry}
\usepackage{graphicx}

\def\beq{\begin{equation}}
\def\eeq#1{\label{#1}\end{equation}}
\def\eeqn{\end{equation}}

\def\beqa{\begin{eqnarray}}
\def\eeqa#1{\label{#1}\end{eqnarray}}
\def\eeqan{\end{eqnarray}}

\let\bar=\overbar

\def\Dslash{\not{\hbox{\kern-4pt $D$}}}
\def\dslash{\not{\hbox{\kern-2pt $\del$}}}

\def\msb{{\bar{\ssstyle M \kern -1pt S}}}

\def\Title#1{\begin{center} {\Large {\bf #1} } \end{center}}
\def\Author#1{\begin{center} {\normalsize {\sc #1} } \end{center}}
\def\Institution#1{\begin{center} {\normalsize {\it #1} } \end{center}}
\def\Abstract#1{\noindent {\normalsize {\bf Abstract:} {\normalfont #1}}}
\def\Conference{\vspace{4mm}\begin{raggedright} {\normalsize {\it Talk presented at the 2019 Meeting of the Division of Particles and Fields of the American Physical Society (DPF2019), July 29--August 2, 2019, Northeastern University, Boston, C1907293.} } \end{raggedright}\vspace{4mm}}

\begin{document}

%
%

\Title{CAPP-8TB: Search for Axion Dark Matter in a Mass Range of 6.62 to 7.04 $\mu$eV}

\Author{Soohyung Lee$^{1}$, Saebyeok Ahn$^{1,2}$, Jihoon Choi$^{1}$, Byeong Rok Ko$^{1}$, and Yannis K. Semertzidis$^{1,2}$}

\Institution{$^{1}$Center for Axion and Precision Physics Research\\Institute for Basic Science\\ 193, Munji-ro, Yuseong-gu, 34051 Daejeon, Republic of Korea}
\Institution{$^{2}$Department of Physics\\Korea Advanced Institute of Science and Technology\\ 291, Daehak-ro, Yuseong-gu, 34141 Daejeon, Republic of Korea}

\Abstract{The axion is a hypothetical particle proposed to solve the strong $CP$ problem, and also a candidate for dark matter. This non-relativistic particle in the galactic halo can be converted into a photon under a strong magnetic field and detected with a microwave resonant cavity. Relying on this detection method, many experiments have excluded some mass regions with certain sensitivities in terms of axion-photon coupling ($g_{a\gamma\gamma}$) for decades, but no axion dark matter has been discovered to date. CAPP-8TB is another axion haloscope experiment at IBS/CAPP designed to search for the axion in a mass range of 6.62 to 7.04\,$\mu$eV. The experiment aims for the most sensitive axion dark matter search in this particular mass range with its first-phase sensitivity reaching the QCD axion band. In this presentation, we discuss the overview of the experiment, and present the first result. We also discuss an upgrade of the experiment to achieve higher sensitivity.}

\Conference

%
%

\begin{figure}[ht]
\centerline{\includegraphics[width=1.\textwidth]{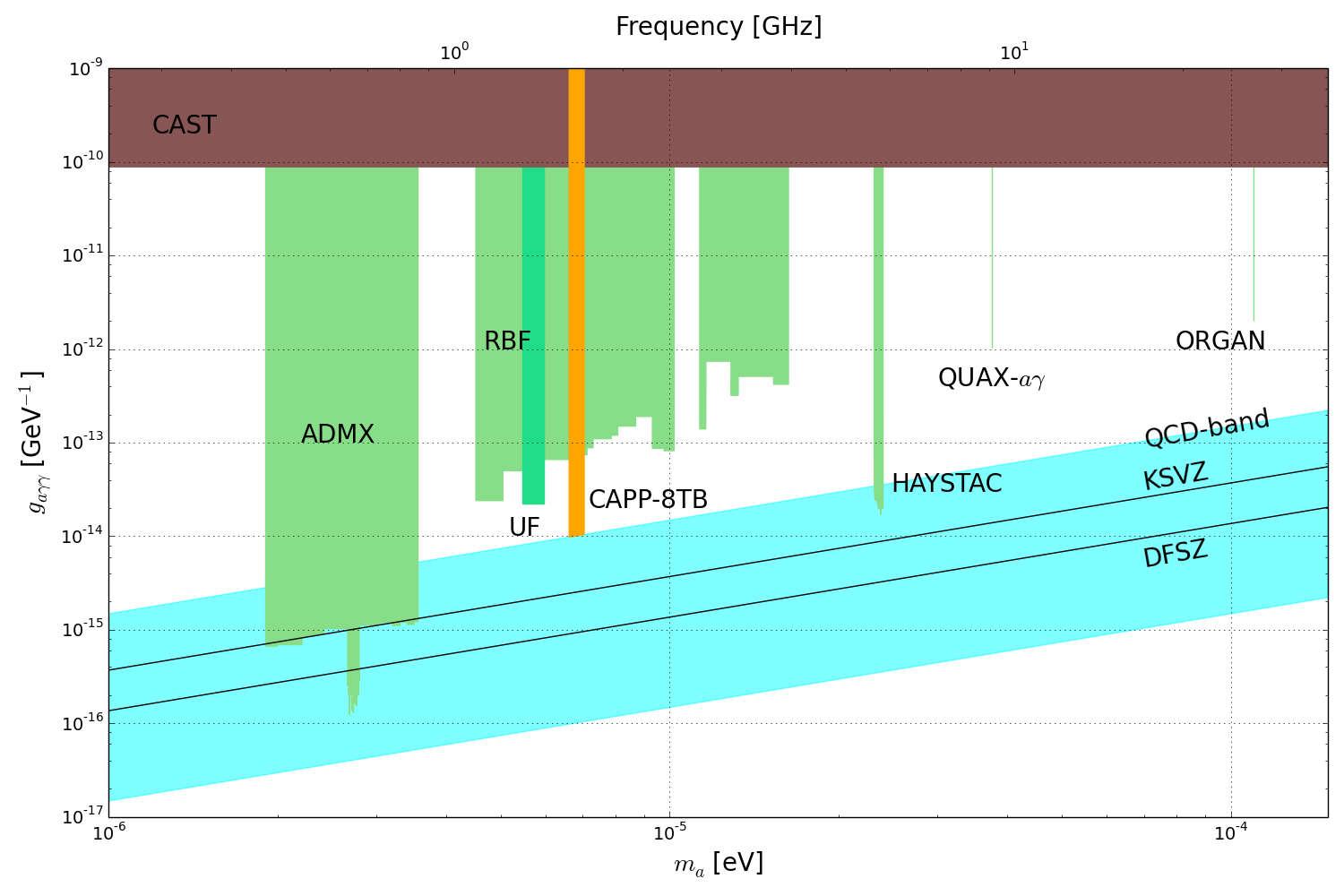}}
\caption{Excluded axion mass and axion-photon coupling by axion haloscope experiments \cite{PhysRevD_64_092003_2001,AstroJLett_571_L27_2002,PhysRevLett_104_041301_2010,PhysRevLett_120_151301_2018,PhysRevD_40_3153_2001,PhysRevD_42_1297_1990,PhysRevLett_118_061302_2017,PhysRevD_97_092001_2018,PhysDarkUniv_18_67_2017,PhysRevD_99_101101_2019} over decades. KSVZ \cite{PhysRevLett_43_103_1979,NuclPhys_B166_493_1980} and DFSZ \cite{PhysLett_104B_199_1981,SovJNuclPhys_31_260_1980} models, and QCD axion band \cite{PhysRevD_52_3132_1995} are shown together. Prospect of the CAPP-8TB experiment is also shown.}\label{fig:current_exclusion}
\end{figure}

The axion \cite{PhysRevLett_40_223_1978,PhysRevLett_40_279_1978} is not only the pseudo-Nambu-Goldstone boson of spontaneously broken global Peccei-Quinn symmetry \cite{PhysRevLett_38_1440_1977} to solve the strong $CP$ problem in quantum chromodynamics (QCD) but also an attractive candidate for a cold dark matter. Over decades, many experiments based on P. Sikivie's method \cite{PhysRevLett_51_1415_1983} attempted to search axion dark matter, however, the axion dark matter has not been found to date. ADMX \cite{PhysRevD_64_092003_2001,AstroJLett_571_L27_2002,PhysRevLett_104_041301_2010,PhysRevLett_120_151301_2018} searched mass ranges of 1.9 -- 3.53\,$\mu$eV for KSVZ \cite{PhysRevLett_43_103_1979,NuclPhys_B166_493_1980} and 2.66 -- 2.81\,$\mu$eV for DFSZ \cite{PhysLett_104B_199_1981,SovJNuclPhys_31_260_1980}, respectively. Rochester-Brookhaven-FNAL \cite{PhysRevD_40_3153_2001} seek a mass range of 4.51 -- 16.25\,$\mu$eV, and University of Florida \cite{PhysRevD_42_1297_1990} searched a mass range of 5.4 -- 5.9\,$\mu$eV. HAYSTAC \cite{PhysRevLett_118_061302_2017,PhysRevD_97_092001_2018} scanned a mass region of 23.15 -- 24.0\,$\mu$eV, ORGAN \cite{PhysDarkUniv_18_67_2017} excluded a mass of 110\,$\mu$eV with a span of 2.5\,neV, and QUAX-$a\gamma$ \cite{PhysRevD_99_101101_2019} explored a mass range of 0.2\,neV around 37.5\,$\mu$eV. The current excluded axion mass and axion-photon coupling is shown in Figure \ref{fig:current_exclusion}. The CAPP-8TB is an axion haloscope with a microwave resonant cavity to search for the axion within a mass range of 6.62 to 7.04\,$\mu$eV. 

The detected axion conversion power in a microwave resonant cavity under a magnetic field is
\begin{eqnarray}
P_{a} & \propto & B^{2}VC_{nlm}Q_{L}
\end{eqnarray}
where $B$ is an external magnetic field, $V$ is a cavity volume, $C_{nlm}$ is a cavity form factor of TM$_{nlm}$ mode, $Q_{L}$ is a loaded quality factor of a cavity. 

\begin{figure}[ht]
\centerline{
\includegraphics[width=0.5\textwidth]{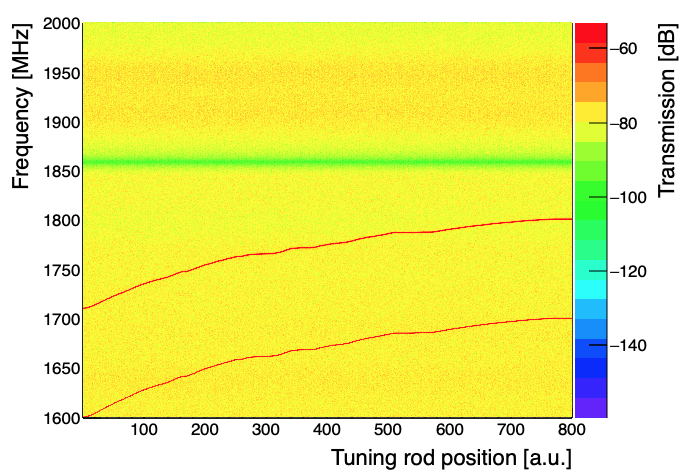}
\includegraphics[width=0.5\textwidth]{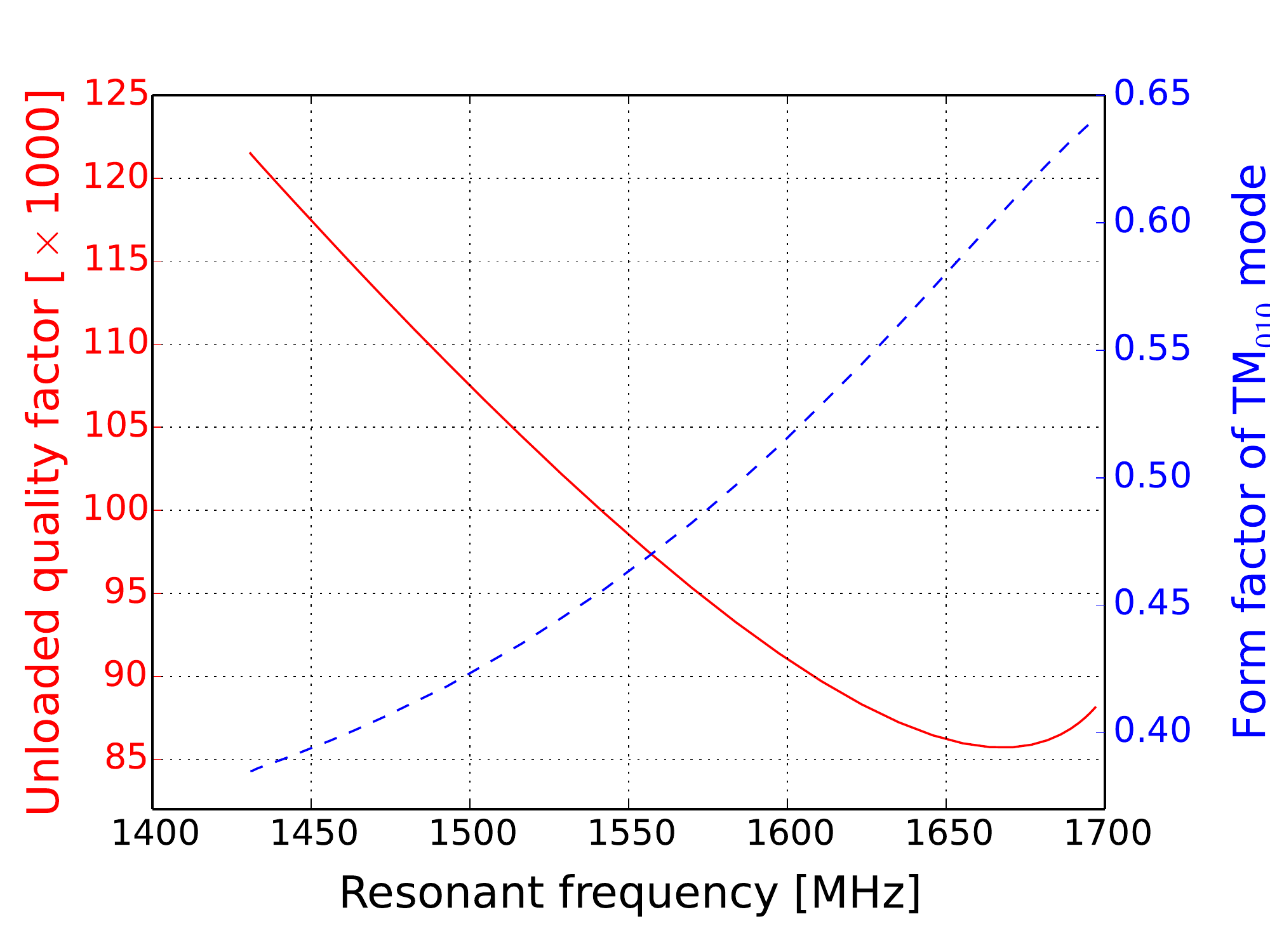}
}
\caption{Left: Measured resonant frequency as the tuning rod rotates. The lower red line which varies from 1600\,MHz to 1700\,MHz is TM$_{010}$ mode. Right: Unloaded quality factor (solid) and form factor of TM$_{010}$ mode (dashed) of the cavity as a function of resonant frequency of the cavity.}\label{fig:cavity}
\end{figure}

In the experiment, a resonant cavity with inner diameter of 134\,mm and inner height of 236\,mm made of copper is used as a detector, therefore, the volume of the cavity is about 3.5 liters. With a dielectric tuning rod, the resonant frequency of the cavity can vary from roughly 1.4 to 1.7\,GHz, and we choose the frequency region to scan in the experiment as 1.6 to 1.7\,GHz which is optimized considering other parameters described later. A critical issue in experiments using resonant cavity is known as \emph{mode-crossing} that the target mode is overlapped with an other mode such as a TE mode. Since there is no way to precisely measure properties of a cavity in this overlapped region, it loses the sensitivity. In simulation studies with our cavity, TM$_{010}$ mode is not overlapped with any other modes, and we also confirm the simulation results from measurements of resonant frequencies as shown in Figure \ref{fig:cavity} (left). Therefore, the experiment will fill full range of frequency range as designed. To maximize the axion conversion power, therefore, the sensitivity of the experiment, we employ TM$_{010}$ mode, and its form factor is found from simulations to be greater than 0.5 for the frequency region of interest as shown in Figure \ref{fig:cavity} (right). Unloaded quality factor is also obtained from simulations as shown in Figure \ref{fig:cavity} (right). The cavity is located in a superconducting magnet which gives a strong magnetic field of 8\,T. Considering the cavity volume, the average magnetic field inside the cavity is estimated as 7.3\,T.

Since the axion power is expected extremely small, spectrum data in the frequency domain has to be averaged properly in an integration time. From the radiometer equation, the signal-to-noise ratio (SNR) is given as
\begin{eqnarray}
\textrm{SNR}\propto\frac{P_{a}}{T_{\textrm{sys.}}}\sqrt{\Delta t}
\end{eqnarray}
where $T_{\textrm{sys.}}$ is a system noise temperature, and $\Delta t$ is an integration time of power spectra. To make the weak axion signal visible, a more integration time and a lower system noise temperature is necessary. The system noise temperature is contributed by a thermal noise of the cavity and intrinsic noise of radio-frequency (RF) components of a receiver chain. To minimize the thermal noise from the cavity, we employ a dilution refrigerator which has base temperature of 10\,mK. We measure the cavity temperature less than 40\,mK, and the tuning rod temperature less than 150\,mK.

\begin{figure}[ht]
\centerline{\includegraphics[width=1.\textwidth]{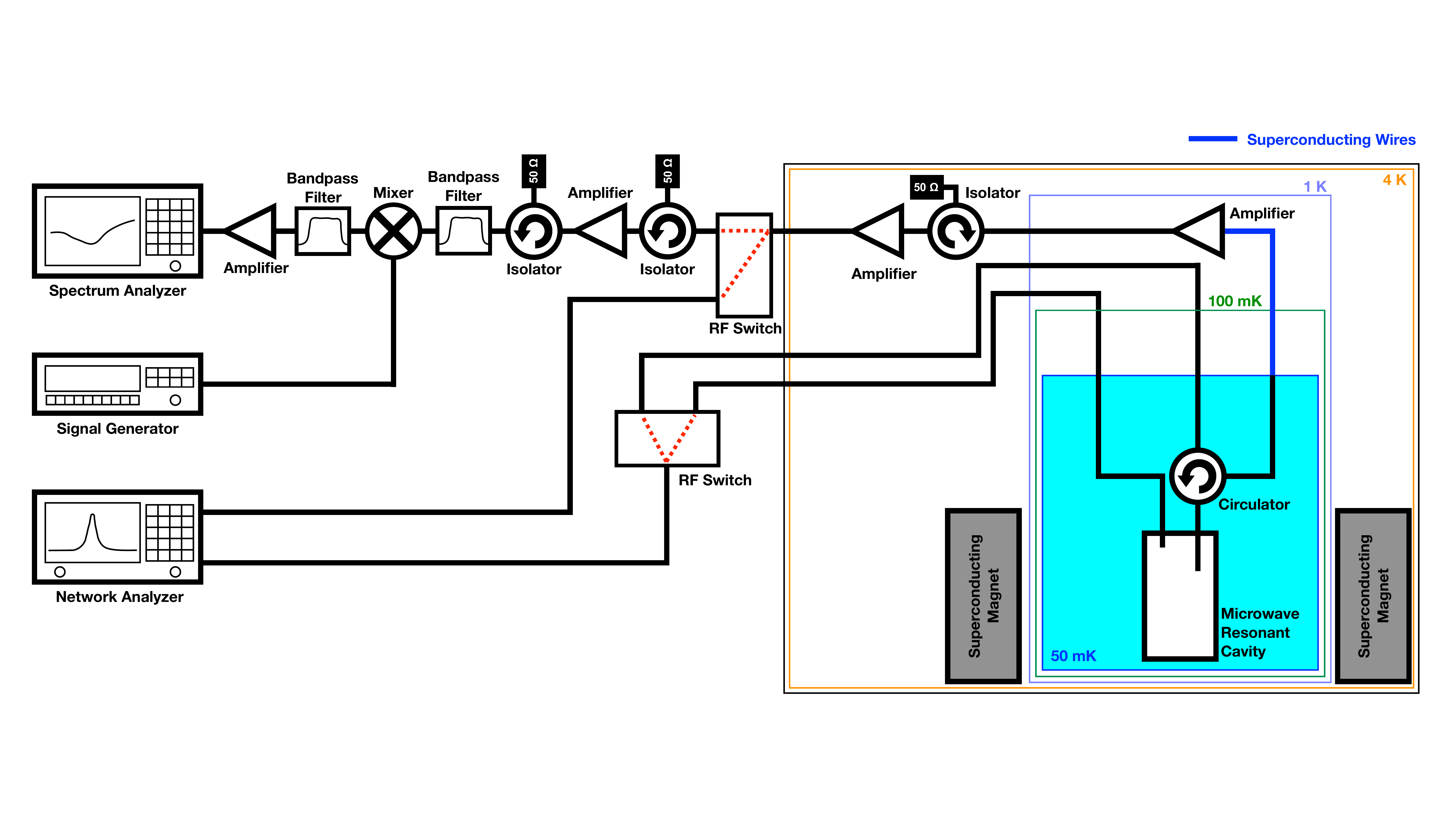}}
\caption{Overview of the CAPP-8TB experiment. The resonant cavity is located at 50\,mK stage, the first amplifier is at 1\,K, and later components are located at higher temperature stages. The magnet is thermally linked to 4\,K stage.}\label{fig:receiver}
\end{figure}

The receiver chain of the experiment is shown in Figure \ref{fig:receiver}. Power spectra from the experiment picked up by the resonant cavity flows to a spectrum analyzer along a series of RF components. To bring the weak power spectra to visible level, we put four amplifiers in the chain: two low-noise amplifiers based on HEMT (high electron mobility transistor) at cryogenic temperatures and two conventional amplifiers at room temperature. The RF power is down-converted to lower frequency by using mixer in front of the last amplifier at room temperature. To block possible resonances from reflections at the amplifiers, we add isolators in front of the amplifiers. In addition, a network analyzer is used to measure the properties of the cavity such as a resonant frequency, a quality factor, and a coupling coefficient to an antenna.

The noise power in a cascade RF chain is given by
\begin{eqnarray}
\label{eq:cascade}
P_{N} & = & k_{B}B\left(T_{\textrm{cavity}} + T_{\textrm{1st amp.}} + \frac{T_{\textrm{2nd amp.}}}{G_{\textrm{1st amp.}}} + \cdots\right)\frac{G}{L}
\end{eqnarray}
where $B$ is a resolution bandwidth, $T_{\textrm{cavity}}$ is a cavity noise temperature, $T_{\textrm{1st amp.}}$ and $T_{\textrm{2nd amp.}}$ are noise temperatures of the first and second amplifiers, respectively, $G_{\textrm{1st amp.}}$ is a gain of the first amplifier, and $G$ is a total system gain, and $L$ is a total attenuation of the system. Since the gain of our first amplifier is about 33\,dB, the third and later terms in parenthesis of Eq. \ref{eq:cascade} are negligible, therefore, the major contribution to the noise power from the receiver chain is the noise temperature of the first amplifier in the chain. We employ one of the best low-noise amplifiers in the market, and the noise temperature of the amplifier is turned out to be less than 1\,K.

To tune the resonant frequency of TM$_{010}$ mode, we employ a pure dielectric tuning rod of Al$_{2}$O$_{3}$, also known as alumina. We also tune the antenna to maintain the coupling coefficient at desired level. Those tunings are driven by stepping motors located outside the refrigerator through series of driving shafts. Since the shafts are thermally linked to different stages in the refrigerator, they should not transmit heats from one to another. To prevent any undesired heat penetrations, the main part of the shafts are made of carbon fiber reinforced polymer, and we have confirmed that the shafts does not transmit any significant heats.

The system including the resonant frequency and coupling coefficient of antenna is controlled by a home-grown data acquisition software, CULDAQ \cite{JPhysConfSer_898_032035_2017}. The software combines the measured power spectra and other data such as cavity response, physical temperatures, and magnetic fields, and store in a ROOT \cite{ROOT} format for a convenience of data analyses.

Through commissioning runs, we have confirmed that the whole system works flawlessly, and obtained reference data such as total gain, noise temperature, and quality factor. In the data taking, we tune the resonant frequency by 20\,kHz, and the tuning typically takes a minute including resonant frequency and antenna coupling tuning and calibrations.

The experiment is currently taking data from 1.60\,GHz. In this run, it targets to touch QCD axion band, and the operation time to scan 100\,MHz for the sensitivity is expected around 3 months. After finishing the data analysis, we will upgrade the system with a quantum-limited noise amplifier, and aim for near KSVZ sensitivity in the same frequency range.

\section*{Acknowledgements}
This work was supported by IBS-R017-D1-2019-a00.


\begin{thebibliography}{99}


\bibitem{PhysRevLett_40_223_1978} S. Weinberg, Phys. Rev. Lett., {\bf 40}, 223 (1978).
\bibitem{PhysRevLett_40_279_1978} F. Wilczek, Phys. Rev. Lett., {\bf 40}, 279 (1978).
\bibitem{PhysRevLett_38_1440_1977} R. D. Peccei and H. R. Quinn, Phys. Rev. Lett., {\bf 38}, 1440 (1977).
\bibitem{PhysRevLett_51_1415_1983} P. Sikivie, Phys. Rev. Lett. {\bf51}, 1415 (1983) doi:10.1103/PhysRevLett.51.1415.
\bibitem{PhysRevD_64_092003_2001} S. Asztalos {\it et al.} (ADMX collaboration), Phys. Rev. D {\bf64}, 092003 (2001) doi:10.1103/PhysRevD.64.092003.
\bibitem{AstroJLett_571_L27_2002} S. J. Asztalos {\it et al.} (ADMX collaboration), Astrophys. J. Lett. {\bf571}, L27 (2002) doi:10.1086/341130.
\bibitem{PhysRevLett_104_041301_2010} S. J. Asztalos {\it et al.} (ADMX collaboration), Phys. Rev. Lett. {\bf104}, 041301 (2010) doi:10.1103/PhysRevLett.104.041301.
\bibitem{PhysRevLett_120_151301_2018} N. Du {\it et al.} (ADMX collaboration), Phys. Rev. Lett. {\bf120}, 151301 (2018) doi:10.1103/PhysRevLett.120.151301.
\bibitem{PhysRevLett_43_103_1979} J. E. Kim, Phys. Rev. Lett. {\bf43}, 103 (1979) doi:10.1103/PhysRevLett.43.103.
\bibitem{NuclPhys_B166_493_1980} M. A. Shifman, A. I. Vainshtein, and V. I. Zakharov, Nucl. Phys. {\bf B166}, 493 (1980)
\bibitem{PhysLett_104B_199_1981} M. Dine, W. Fischler, and M. Srednicki, Phys. Lett. {\bf104B}, 199 (1981) doi:10.1016/0370-2693(81)90590-6.
\bibitem{SovJNuclPhys_31_260_1980} A. R. Zhitnitsky, Sov. J. Nucl. Phys. {\bf31}, 260 (1980).
\bibitem{PhysRevD_40_3153_2001} W. U. Wuensch {\it et al.}, Phys. Rev. D {\bf40}, 3153 (2001) doi:10.1103/PhysRevD.40.3153.
\bibitem{PhysRevD_42_1297_1990} C. Hagmann, P. Sikivie, N. S. Sullivan, and D. B. Tanner, Phys. Rev. D {\bf42}, 1297 (1990) doi:10.1103/PhysRevD.42.1297.
\bibitem{PhysRevLett_118_061302_2017} B. M. Brubaker {\it et al.} (HAYSTAC collaboration), Phys. Rev. Lett. {\bf118}, 061302 (2017) doi:10.1103/PhysRevLett.118.061302.
\bibitem{PhysRevD_97_092001_2018} L. Zhong {\it et al.} (HAYSTAC collaboration), Phys. Rev. D {\bf97}, 092001 (2018) doi:10.1103/PhysRevD.97.092001.
\bibitem{PhysDarkUniv_18_67_2017} B. T. McAllister {\it et al.}, Phys. Dark Universe {\bf18}, 67 (2017) doi:10.1016/j.dark.2017.09.010.
\bibitem{PhysRevD_99_101101_2019} D. Alesini {\it et al.}, Phys. Rev. D {\bf99}, 101101(R) (2019) doi:10.1103/PhysRevD.99.101101.
\bibitem{PhysRevD_52_3132_1995} S. L. Cheng, C. Q. Geng, and W. -T. Ni, Phys. Rev. D {\bf52}, 3132 (1995) doi:10.1103/PhysRevD.52.3132.
\bibitem{JPhysConfSer_898_032035_2017} S. Lee, J. Phys.: Conf. Ser., {\bf 898}, 032035 (2017).
\bibitem{ROOT} R. Brun and F. Rademakers, Nucl. Instrum. Meth. Phys. Res. A {\bf389}, 81 (1997).

\end{thebibliography}
\end{document}